\documentclass[preprint,prl,showpacs]{revtex4}
\usepackage{amsmath}

\input{tcilatex}

\begin{document}

\noindent \textbf{Helix-coil Transition in Closed Circular DNA} \vskip1 cm

\noindent V. F. Morozov$^{*,+}$, Shura Hayryan$^*$, E. Sh. Mamasakhlisov$%
^{*,+}$, A. V. Grigoryan$^+$,\newline
 A. V. Badasyan$^+$,and Chin-Kun Hu$^{*}$

\noindent $^*$Institute of Physics, Academia Sinica, Nankang, Taipei 11529,
Taiwan, $^+$Department of Molecular Physics, Yerevan State University, 1
Al.Manougian Str., Yerevan 375025, Armenia \vskip 1 cm

\noindent{ABSTRACT}{~~~} A simplified model for the closed circular DNA
(ccDNA) is proposed to describe some specific features of the helix-coil
transition in such molecule. The Hamiltonian of ccDNA is related to the one
introduced earlier for the linear DNA. The basic assumption is that the
reduced energy of the hydrogen bond is not constant through the transition
process but depends effectively on the fraction of already broken bonds. A
transformation formula is obtained which relates the temperature of ccDNA at
a given degree of helicity during the transition to the temperature of the
corresponding linear chain at the same degree of helicity. The formula
provides a simple method to calculate the melting curve for the ccDNA from
the experimental melting curve of the linear DNA with the same nucleotide
sequence. \vskip 1.0 cm


\section{Introduction}

In its biological ``native'' state, a DNA molecule has a form of well-known
right-handed double helix (Crick and Watson, 1954), wherein two
heteropolymer chains are wound around each other. The double helical
structure is believed to be the structure of minimum free energy, under the
normal physiological conditions. It is stabilized by many factors, among
which the most essentials are hydrogen bonds between complementary
nitrogenous bases (A-T and G-C) on opposite chains and the interactions
between neighboring base pairs along the chain (stacking of base pairs),
which have a hydrophobic nature. When the environmental conditions are
changed, two chains can separate from each other along some parts of the
molecule, giving rise to the loops, bordered by helical regions. The double
helix can also unwind completely and split into two separate chains. Thus
the whole molecule or its parts undergo the transition from the state of
energetically favorable high-ordered helical structure into the state of
disordered coil with large entropy. This process is known as helix-coil
transition. Alternative names are ``melting'' or ``denaturation''. The
helix-coil transition can be caused by many factors. In the living cell it
is mediated by specific protein molecules, in \textit{in vitro} experiments
it may be realized by changing the temperature, chemical composition, the
concentration of salts, etc. of the DNA solution.

The helix-coil transition in DNA has been a subject of very intensive
theoretical investigations since 1960s (Poland and Scheraga, 1970; Flory,
1969; Volkenshtein, 1977; Vedenov et al., 1972; Wartell and Benight, 1985;
Wada and Suyama., 1986; Grosberg and Khokhlov, 1994); for the foundations of
theory, see e.g. (Poland and Scheraga, 1970; Flory, 1969; Volkenshtein,
1977) ; for reviews of earlier research works, see e.g. (Vedenov et al.,
1972; Wartell and Benight, 1985; Wada and Suyama., 1986); for review of more
recent developments, see e.g. (Grosberg and Khokhlov, 1994) Here we make
just several very short remarks on the background of the helix-coil
transition.

The most interesting feature of the helix-coil transition in DNA is its
cooperativity which is a manifestation of long-range interactions along the
chain. Many factors can influence the degree of the cooperativity. It
depends, for example, on the base sequence of DNA, on the chain length,
ionic strength of the solution, etc.

The usual way for describing the helix-coil transition is to find the
dependence of the \textit{degree of helicity}, $\theta$, on the external
parameter (e.g. the temperature); $\theta$ is defined as the average
fraction of the bounded pairs: $\theta={<n>}/{N}$, where $N$ is the total
number of base pairs and $<n>$ is the average number of the bounded pairs.
The graph of this dependence is called a ``melting curve''. The temperature
at which $\theta=1/2$ is called melting point ($T_m$) and is one of the
characteristics of the melting process. Another quantitative characteristic
is the ``melting interval'' or the width of melting $\Delta T=[\frac{%
\partial \theta}{\partial T}]^{-1}$, where the derivative is taken at the
point $T_m$. $\Delta T$ is an actual measure of the cooperativity. The
smaller is the melting interval the more cooperative is the transition.

Traditionally the theoretical models for helix-coil transition are based on
the assumption that every base pair can be in two possible states:
hydrogen-bonded (helix) or open (coil). This makes it convenient to use the
Ising model for ferromagnets as a basic tool to describe the helix-coil
transition. Usually four parameters are introduced into models: (1) The
helix stability parameter $s$, called also the equilibrium constant of helix
growth. It corresponds to the statistical weight of the helical base pair,
which follows a helical one. (2) The cooperativity parameter $\sigma$ (or
helix initialization constant), which corresponds to the statistical weight
of a bonded pair at the junction of the helical and coil regions of the
chain. The actual values of these two parameters are averaged over all
conformations of the molecule or over the ensemble of identical molecules.
(3) The loop-weighting factor $m^{-\alpha}$, where $m$ is the number of
bases in the loop, and $\alpha=1.5-1.7$ is called \textit{%
Jackobson-Stokmayer exponent} (Jacobson and Stockmayer, 1950). The meaning
of this factor is that the entropy of the melted loop of $m$ base pairs is
different from the entropy of $m$ open pairs at the free ends of the
molecule. In other words, the entropy is not additive inside the loop
(Poland and Scheraga, 1966) (4) Another factor, which affects the nature of
helix-coil transition is called \textit{dissociation equilibrium constant}.
It corresponds to the process, when the last bond opens, and the molecule
splits into two fully separated strands. Note that the classical Ising model
does not contain last two factors and must be modified correspondingly.

An even more topologically complex object is the closed circular DNA
molecule, which typically occurs in some simple biological systems
(plasmids, viruses) and in cytoplasm of the animal cells (Dulbecco and Vogt,
1963; Weil and Vinograd, 1963). In this form the double-helical DNA is
twisted in such a way that the first monomer of the sugar-phosphate backbone
of each chain is covalently linked to the last one to make up a closed
circle (Fig. 1). The topologically similar structures may occur also in the
cells of higher animals in which some DNA molecules are wound around the
specific protein structures (histones). Double helical closed circular DNA
(ccDNA) typically exists in the conformation of so called supercoil which
represents a sort of interwound structure. The degree of supercoiling as
well as the topological state of supercoil are described by a linking number
(Vologodskii, 1992). Many papers on experimental and theoretical
investigations of ccDNA have been published (Marko and Siggia, 1995 and
references therein). One of the interesting problems is the helix-coil
transition in these molecules. Because of the specific topological
restraints some features of the helix-coil transition in ccDNA differ from
those of linear DNA. For example, experiments show that the melting process
of a ccDNA begins at a lower temperature than that in a linear DNA, and
completes at considerably higher temperatures (Vinograd et al., 1968). In
(Laiken, 1973; Anshelevich et al., 1979; Vologodskii et al., 1979; Benham,
1979; Benham, 1980; Vologodskii and Frank-Kamenetskii, 1981; Belintsev, B.
N. and A.V. Gagua. 1989) a mean field theory for melting of the ccDNA is
developed. The general assumption in these works is that the total energy of
ccDNA consists of two parts: the term corresponding to the ordinary DNA, and
the term corresponding to the fact that the molecule under description is a
closed circular system (superhelix term).

In (Hayryan et al., 1990; Hayryan et al., 1995) some of us have developed a
microscopical theory of helix-coil transition in polypeptide chains which
doesn't include averaged phenomenological parameters like $s$ and $\sigma$
and is based on the molecular characteristics of the chain. A many-particle
Potts-like model was used instead of the two-state Ising model. Since the
model was able to reproduce qualitatively many important characteristics of
the helix-coil transition in polypeptides, we were encouraged to use a
similar approach to the linear DNA molecule. It appeared that the secular
equation of the DNA Hamiltonian may be written in exactly the same form as
for polypeptide chain in which the characteristic length of hydrogen bond is
replaced by some characteristic length of DNA molecule. The model was
published in (Morozov et al., 2000) and will be reviewed briefly in Sec. II
of the present paper.

The purpose of the present work is to continue the microscopical approach by
applying it to the melting process of ccDNA, using an Hamiltonian approach,
similar to one which was used for polypeptides and linear DNA. We are going
to show that the model of ccDNA can be reduced to the model of linear DNA
(which was in its turn reduced to the polypeptide model with appropriate
redefinition of parameters). We do not consider explicitly the topological
restraints, nor the superhelic structure of ccDNA. Instead we impose some
specific conditions on the mechanism of breaking of the hydrogen bonds. We
obtain a transformation formula which relates the temperature of ccDNA for a
given degree of helicity to the temperature of the linear DNA for the same
degree of helicity. Using this transformation formula, we suggest a simple
method to obtain the melting curves for the ccDNA from the melting curves of
the corresponding linear DNA.

This paper is organized as follows: In Sec. II we briefly review the model
of linear DNA. In Sec. III we introduce the model of ccDNA. In Sec. IV the
melting curve for ccDNA is related to melting curve for linear DNA. Some
problems for further studies are discussed in Sec. V.

\section{The Model of the Linear DNA}

Consider a double-chain DNA molecule which consists of solely one kind of
complementary nitrogen base pairs, either A-T or G-C, which are displaced
randomly along the chain. Let the monomers in each chain be enumerated $%
0,1,2,...,N$. In this case one can assume that the inter-chain hydrogen
bonds are formed only between the bases with the same order number because
the probability of mismatching of the bases is extremely small due to the
randomness of the sequence.

We construct the model of such a system as follows (Morozov et al., 2000) .
To each repeated unit $i$ of one chain of the double helix a vector $\vec
a_i $ is assigned. Similarly, the vector $\vec b_i$ is assigned to the $i$%
-th unit of the opposite chain. One can consider these vectors as directed
along the line connecting two adjacent sugar rings. We also assign a vector $%
\vec d_i$ to each complementary pair of nitrogen bases. The magnitude of
this vector is not important. We will just assume that all$\vec d_i$ emanate
always from the same chain. When the corresponding complementary pair is in
the helical conformation, the vector $\vec d_i$ connects the ends of vectors
$\vec a_i$ and $\vec b_i$. For the sake of simplicity, we assume that the
first (number $0$) complementary pair is always in the helical conformation
and the bases are connected by the vector $\vec d_0$. Figure 2 shows the
scheme of hydrogen bond formation.

When the $i$-th complementary pair is also in the helical conformation then
we assume that the loop $0-(i-1)$ is formed. Geometrically this means that

\begin{equation}  \label{d1}
-\vec d_0+\sum_{k=1}^i\vec a_k+\vec d_i-\sum_{k=1}^i\vec b_k=0,
\end{equation}
Let $\vec \gamma_k=-\vec d_{k-1}+\vec a_k-\vec b_k+\vec d_k$, then Eq. (\ref%
{d1}) may be rewritten as

\begin{equation}  \label{d2}
\sum_{k=1}^i\vec \gamma_k=0.
\end{equation}
Note that if the hydrogen bond in the $i$-th pair is formed then (\ref{d2})
holds true even if there are no other hydrogen bonds in any of the pairs
between the first and the $i$-th pairs.

The Hamiltonian of the chain then reads

\begin{equation}  \label{hamil}
-\beta H=J\sum_{i=1}^N\delta \left( \sum_{k=1}^i\vec \gamma_k,0 \right),
\end{equation}
where $J=U/T$, $\beta =T^{-1}$ and $\delta$ is the Kronecker delta symbol
with $U$ being the energy for hydrogen bond formation in one complementary
pair. To be more precise, the quantity $U$ is the energy of the helical
state of one base pair, i.e., it includes also the stacking of bases
(Morozov et al., 2000). Here an approximation is done that the stacking
interactions are the same for all base pairs throughout the chain. Further
in the text we will use for (\ref{hamil}) the form

\begin{equation}  \label{hamil1}
-\beta H=J\sum_{i=1}^N\delta_{1}^{(i)},
\end{equation}
where the notation $\delta_{1}^{(i)} =\delta \left( \sum_{k=1}^i\vec
\gamma_k,0 \right)$ is introduced.

We should mention an important feature of the Hamiltonian (\ref{hamil}).
Though the first sum is extended over the number of base pairs, it doesn't
mean that the contributions from different base pairs are independent. The
term in the brackets shows that the state of $i$th pair, hence its
contribution, depends on the states of all previous $(i-1)$ pairs. Thus, the
cooperative interdependence of successive linked base pairs is implicitly
included through real geometrical restrictions.

For infinitely long chain one has the following equations for the partition
function and the free energy, respectively:

\begin{equation}  \label{Z}
Z=P_0^{-N},
\end{equation}

\begin{equation}  \label{F}
F=TlnP_0,
\end{equation}
where $P_0$ is the nearest to zero root of the secular equation for
Hamiltonian (\ref{hamil1}) (Hayryan et al., 1990; Morozov et al., 2000)

\begin{equation}  \label{sec}
\sum_{m=1}^{\infty}P^m\varphi (m)=\frac{1}{V},
\end{equation}
where $V=e^J-1$ and

\begin{equation}  \label{phi}
\varphi (m)=Q^{-m}\sum_{\gamma_{1}}\sum_{\gamma_{2}} \cdots
\sum_{\gamma_{m}}\delta_{i}^{(m)}.
\end{equation}
Here $Q$ is the number of conformations of the repeated unit and bears the
same meaning as in the case of polypeptide chain (Hayryan et al., 1990;
Hayryan et al., 1995) . Namely,

\begin{equation}  \label{q}
Q=\frac{\mathrm{Partition~ function ~of~ one ~repeated~ unit}} {\mathrm{%
Partition~ function~ of~ one~ repeated~ unit~ in ~helical~ structure}}.
\end{equation}
The meaning of the function $\varphi(m)$ may be interpreted as the ratio of
the partition function of the loop of $m$ units to that of the same chain
without loops.

The secular equation (\ref{sec}) contains two microscopical quantities: $(i)
$ the temperature parameter $V$, which contains the energy of inter-chain
hydrogen bonding, $(ii) \varphi(m)$ which represents the relative
statistical weight of the loop of length $m$. Both quantities can, in
principle, be measured in experiments or be calculated by other independent
methods. The function $\varphi(m)$ behaves differently for the small and
large values of $m$ (see Morozov et al., 2000 for details). It has been
shown (Morozov et al., 2000) that the partition function for the Hamiltonian
(\ref{hamil1}) of the linear DNA can be reduced to the Hamiltonian of the
generalized polypeptide model (Hayryan et al., 1990; Hayryan etal., 1995)

\begin{equation}  \label{hamil2}
-\beta H=J\sum_{i=1}^N\prod\nolimits_{k=0}^{\Delta -1}\delta (\gamma_{i+k},1)
\end{equation}
with the secular equation

\begin{equation}  \label{sec1}
\lambda^{\Delta -1} (\lambda -e^{J})(\lambda -Q)=(e^{J}-1)(Q-1),
\end{equation}
where $\Delta$ is the number of amino acid residues embraced by one
intramolecular hydrogen bond for polypeptides, and $\Delta=$ the persistent
length, for DNA.

In a similar way, it is possible to reduce the Hamiltonian of ccDNA also to
the Hamiltonian of generalized polypeptide model. We are going to show this
in the next section.

\section{The Model of Closed Circular DNA}

One can see easily that it is impossible to separate two chains of ccDNA
completely, without breaking chemical bonds. Suppose at some part of the
molecule some hydrogen bonds have been broken and a loop has formed. Then
further growing of the loop or the formation of other loops becomes more and
more difficult because there is no entropy gain to compensate the energetic
losses. This means that the denaturing rate at each point of the molecule
will depend on the conformation of the whole chain. Thus ccDNA is a system
in which the state of repeated unit depends on the state of the whole
molecule. Starting from this general observation and from the above
described model of linear DNA, we construct the Hamiltonian for the ccDNA as
follows.

We assume that in the Hamiltonian (\ref{hamil1}) the instantaneous value of
the reduced energy of the hydrogen bonds $J=J(\eta)$ is a function of the
fraction $\eta$ of the broken hydrogen bonds in the molecule

\begin{equation}  \label{p}
\eta=1-\frac{1}{N}\sum_{k=1}^N\delta_{1}^{(k)}.
\end{equation}
Note that $\eta$ is not an averaged quantity. It characterizes the
instantaneous degree of denaturing.

As a first step let us imagine $J(\eta )$ to be linear function of $\eta $

\begin{equation}
J(\eta )=J_{0}+a+b\eta .  \label{J}
\end{equation}%
Here $J_{0}=U/T$, $U$ is the energy of hydrogen bond, $a$ and $b$ are some
coefficients which depend on temperature.

Then the Hamiltonian (\ref{hamil1}) can be written as

\begin{equation}
-\beta H=(J_{0}+a+b)\sum_{i=1}^{N}\delta _{1}^{(i)}-\frac{b}{N}\left[
\sum_{i=1}^{N}\delta _{1}^{(i)}\right] ^{2}  \label{hamcc}
\end{equation}%
%
%
%
%
%
The conformational partition function corresponding to the Hamiltonian (\ref%
{hamcc}) is

\begin{equation}
Z=\sum_{\{\gamma _{k}\}}exp\left( (J_{0}+a+b)\sum_{k=1}^{N}\delta
_{1}^{(k)}\right) \times exp\left( \frac{b}{N}\left[ i\sum_{k=1}^{N}\delta
_{1}^{(k)}\right] ^{2}\right)
\end{equation}%
The imaginary unity in the argument of the second multiplier is introduced
to ensure the positiveness of this term. This equation can be simplified by
using the Hubbard-Stratanovich identities:

\begin{equation}
exp\left( \frac{\varphi ^{2}}{2g}\right) \sim \int_{-\infty }^{\infty
}\left( -\frac{g}{2}x^{2}+\varphi x\right) dx,  \label{hub1}
\end{equation}

Hence

\begin{equation}
Z\propto \int\limits_{0}^{\infty }dxexp\left( -\frac{1}{4}\frac{N}{b}%
x^{2}\right) \times \left[ \sum_{\{\gamma
\}}exp(J_{0}+a+b+ix)\sum_{k=1}^{N}\delta _{1}^{(i)}\right] .
\end{equation}

In (\ref{zcc1}) the expression included in the square brackets is identical
to the partition function $Z_{0}$ of a system with Hamiltonian (\ref{hamil1}%
) in which the reduced energy of the hydrogen bond formation is replaced by

\begin{equation}
J=J_{0}+a+b+ix.  \label{J1}
\end{equation}%
In the thermodynamical limit

\begin{equation}  \label{z0}
Z_{0}\approx\lambda^N,
\end{equation}
where $\lambda$ is called \textit{maximum characteristic Lyapunov exponent}
(Crisanti et al., 1993) for Hamiltonian (\ref{hamil1}) and is a function of $%
J$ described in (\ref{J1}).

At this point we can state that by redefinition of the hydrogen bond energy
the model of ccDNA is reduced to the model of linear DNA which was itself
reduced to the polypeptide model.

Then the partition function for the ccDNA is transformed to

\begin{equation}
Z\approx \int_{0}^{\infty }dxexp\left( N\left( -\frac{1}{4}\frac{x^{2}}{b}%
+ln\lambda \right) \right) .  \label{zcc2}
\end{equation}%
For large $N$ the integral can be evaluated using the saddle point method

\begin{equation}
Z\approx exp\left( N\left( -\frac{1}{4}\frac{x_{0}^{2}}{b}+ln\lambda
_{0}\right) \right) ,  \label{zcc3}
\end{equation}%
where $\lambda _{0}$ is the value of Lyapunov exponent at the saddle point
and is a function of argument $(J_{0}+a+b+ix_{0})$, $x_{0}$ is obtained from
the condition that the element of integration takes its maximum at the
saddle point.

For $\lambda$ we have from Hamiltonian (\ref{hamil1})

\begin{equation}  \label{dldJ}
\frac{\partial ln\lambda}{\partial J}=\frac{1}{NZ}\sum_{\{\gamma_{i}\}}
\sum_{i}\delta_{1}^{(i)}exp\left(J\sum_{i}
\delta_{1}^{(i)}\right)\equiv\theta,
\end{equation}
where $\theta$ is the degree of helicity for the open DNA or the average
fraction of the repeated units in the helical conformation.

The free energy per repeated unit is

\begin{equation}
f=ln\lambda (J)+b\theta ^{2},  \label{freeen}
\end{equation}%
where $J$ from (\ref{J1}) is reduced to

\begin{equation}
J=J_{0}+a-b+2b(1-\theta ).  \label{J2}
\end{equation}%
Thus we see that the helicity degree $\theta $ for the linear chain emerges
in the partition function of circular molecule as a parameter.

\section{Melting Curves for Closed Circular DNA and linear DNA}

From the partition function, we can evaluate the helicity degree of ccDNA

\begin{equation}  \label{curve1}
\Theta =\frac{1}{N}\frac{\partial lnZ}{\partial J_{0}}=\frac {\partial f}{%
\partial J_{0}}.
\end{equation}
Straightforward calculations lead to

\begin{equation}
\Theta =\theta \cdot \left[ 1-2b\frac{\partial \theta }{\partial J_{0}}%
\right] +\left[ 2b\theta \right] \frac{\partial \theta }{\partial J_{0}}%
=\theta .  \label{curve2}
\end{equation}%
Though the equations for the free energy are different for ccDNA and linear
DNA, the dependence of helicity degree on the respective reduced energy of
hydrogen bonds of each model is the same. This means that the degree of
helicity in ccDNA can be obtained from that of linear chain by redefinition
of reduced energy of hydrogen bonding by the formula (\ref{J2}).

From (\ref{J2}) one can easily obtain the relation between the temperatures
corresponding to the equal values of helicity degree for linear and ccDNA.

\begin{equation}
T_{C}=T_{L}\left[ 1+(\alpha -\beta )+2\beta (1-\theta )\right] .
\label{temp}
\end{equation}%
Here $T_{C}$ and $T_{L}$ are temperatures of ccDNA and linear DNA,
respectively, $\alpha =a/J_{0},\beta =b/J_{0}$. Eq. (\ref{temp}) allows to
calculate the denaturing curve for ccDNA from the melting curve of
corresponding linear molecule (Fig.3).

In (Gagua et al., 1981) the experimental differential denaturing curve of
the mixture of linear and ccDNA in the solvent has been presented, which
shows equality of melting temperatures of GC and AT pairs. One can observe
that the transition in the linear molecule occurs within a much more smaller
temperature interval $(\sim 1^{\circ })$ than in ccDNA $(\sim 20^{\circ })$.
Besides, the melting process in ccDNA begins earlier and completes at higher
temperature. Let us try to explain this fact in the framework of the present
model. For the sake of simplicity suppose that the melting of linear DNA
occurs as a pure phase transition (Fig. 4a) and let $\alpha -\beta <0$, $%
\beta >0$. Then the denaturing curve of ccDNA from Eq. (\ref{temp}) will be
represented by a linear function of temperature.

\begin{equation}  \label{denatur}
1-\theta=\frac{T}{T_M}\cdot \frac{1}{2\beta}+\frac{\beta- \alpha-1}{2\beta},
\end{equation}
where $T_M$ is the transition temperature. This equation is true in the
temperature interval from $T_1= T_M(1-(\beta-\alpha))$ to $%
T_2=T_M(1+\alpha+\beta)$. As shown in Fig. 4b, we have

\begin{equation}
\frac{\partial (1-\theta)}{\partial T}=\frac{1}{2\beta T_M},  \label{Dtheta}
\end{equation}
for $T_1 < T < T_2$. This means that while the linear DNA undergoes a sharp
transition at certain temperature $T_M$, the transition in ccDNA has a
finite interval, given by Eq. (\ref{Dtheta}). Thus, within the present model
it is possible to describe qualitatively considerable widening of the
transition interval of ccDNA as compared to the linear molecule as well as
shifting of the left point of the transition interval toward low
temperatures.

Another experimental fact is that the differential melting curve of linear
DNA of higher organisms has very rugged form while the melting curve of
ccDNA is relatively smooth.This result can be explained as follows. It is
widely believed that each peak at the DMC corresponds to the melting of
particular region of DNA. Thus, each of these regions melts as mini-DNA.
Consequently, every peak at the DMC corresponds to the S-like region of the
melting curve, containing inflection point. The temperature transformation,
described by Eq.(\ref{preobT}), makes each of these regions more flat in
analogy with the whole melting curve. It makes the DMC much more smooth. Our
calculations (in agreement with our previous results obtained in the
particular case) show that the curves are smooth, melting begins earlier and
the transition interval is very large. One can see that the calculated
curves possess all three features of ccDNA: the curves are smooth, melting
begins earlier and the transition interval is very large.

\section{Conclusion}

We have constructed the model for the ccDNA using the earlier developed
model for the linear DNA. The connection between two models is defined by
the universal behavior of the order parameter (helicity degree) vs the
reduced energy of formation of hydrogen bonds between complementary pairs.
The parameters $\alpha $, $\beta $ in the expression of $J(\eta )$ are
closely related to the topology and energetics of superhelic structure. In
further work it is necessary to try to establish these relations explicitly.

It would be interesting also to include higher power terms in Eq. (\ref{J})
and evaluate the corresponding transformation formula for the melting
temperatures.

\begin{eqnarray}  \label{high}
T_C=T_L(1+(\alpha-\beta)+2(\beta-\gamma)(1-\theta)+  \notag \\
3(\gamma-\delta)(1-\theta)^2+4(\delta-\epsilon)(1-\theta)^3 \cdots.
\end{eqnarray}
Note that the comparison of the results with the experimental data is done
only qualitatively. We have just predicted the widening of the melting
interval, smoothening of the melting curve and disappearing of some sequence
dependent details.

To compare the theory with the experimental data quantitatively, it is
necessary to carry out experiments on melting of the ccDNA with certain
superhelicity (with certain $\alpha$, $\beta$, $\gamma$ ...) and of the
linear chain under the identical conditions.

\vskip 0.5 cm This work was supported in part by the National Science
Council of the Republic of China (Taiwan) under Contract No. NSC
90-2112-M-001-074. CRDF Foundation under Grant No. AB2-2006, and ISTC under
Grant No. A-301d.

\vskip 1.0 cm \noindent\textbf{REFERENCES}\newline

\noindent Anshelevich, V. V., A. V. Vologodskii, A. V. Lukashin, and M. D.
Frank-Kamenetskii. 1979. Statistical-mechanical treatment of violations of
the double helix in supercoiled DNA. \textit{Biopolymers.} 18: 2733-2744.
\newline

\noindent Belintsev, B. N. and A.V. Gagua. 1989. Why is the process of
melting of supercoiled DNA so noncooperative. {Mol. Biol. } (Russian) 23:
37-43.\newline

\noindent Benham, C. J. 1979. Torsional stress and local denaturation in
supercoiled DNA. \textit{Proc. Natl. Acad. Sci. USA} 76: 3870-3874. \newline

\noindent Benham, C. J. 1980. The Equilibrium statistical mechanics of the
helix-coil transition in torsional stressed DNA. \textit{J. Chem. Phys.} 72:
3633-3639. \newline

\noindent Crick, F. H. C. and J. D. Watson. 1954. The complementary
structure of deoxyribonuleic acid. \textit{Proc. Roy. Soc. (London) A}, 223:
80. \newline

\noindent Crisanti,A., G. Paladin, A. Vulpiani, in \emph{Products of Random
Matrices in Statistical Physics} (Springer-Verlag, Berlin, 1993).\newline

\noindent Dulbecco, R. and M. Vogt. 1963. Evidence of a ring structure of
polyoma virus DNA. \textit{Proc. Natl. Acad. Sci. U.S.A.} 50: 236. \newline

\noindent Flory, P. J. \emph{Statistical Mechanics of Chain Molecules}
(Interscience, New York, 1969). \newline

\noindent Gagua, A. V., B. N. Belintsev, and Yu. L. Lyubchenko. 1981. Effect
of base-pair stability on the melting of superhelical DNA. \textit{Nature}
294: 662-663. \newline

\noindent Grosberg A. Yu. and A. R. Khokhlov, \emph{Statistical Physics of
Macromolecules} (AIP Press, New York, 1994).\newline

\noindent Hayryan, Sh. A, N. S. Ananikian, E. Sh. Mamasakhlisov, V. F.
Morozov. 1990. The helix-coil transition in polypeptides -- A microscopic
approach \textit{Biopolymers.} 30: 357- 367.\newline

\noindent Hayryan, Sh. A., E. Sh. Mamasakhlisov, V.F. Morozov. 1995. The
helix-coil transition in polypeptides --A microscopic approach 2. \textit{%
Biopolymers.}35: 75-84. \newline

\noindent Jacobson, H. and W. Stockmayer. 1950. Intramolecular reaction in
polycondensation. 1. The theory of linear systems. \textit{J. Chem. Phys.}
18: 1600-1606. \newline

\noindent Laiken, N. 1973. Theoretical model for equilibrium behavior of DNA
superhelices. \textit{Biopolymers.} 12: 11-26. \newline

\noindent Marko, J. F. and E. D. Siggia. 1995. Statistical mechanics of
supercoiled DNA. \textit{Phys. Rev. E} 52: 2912-2938. \newline

\noindent Morozov V. F., E. Sh. Mamasakhlisov, and M. S. Shahinyan. 1998.
\textit{J. Contemporary Physics.} 33: 195. \newline

\noindent Morozov V. F., E. Sh. Mamasakhlisov, Sh. Hayryan, and C.-K. Hu.
2000. Microscopical approach to the helix-coil transition in DNA. \textit{%
Physica A.} 281: 51-59.\newline

\noindent Poland, D. and H. A. Scheraga. 1966. Occurrence of a phase
transition in nucleic acid models. \textit{J. Chem. Phys.} 45: 1464. \newline

\noindent Poland D. C. and H. A. Scheraga, \emph{The Theory of Helix-Coil
Transition} (Acad. Press, New York, 1970).\newline

\noindent Vedenov, A. A., A. M. Dykhne and M. D. Frank-Kamenetsii. 1972.
Helix-coil transition in DNA. Sov. \textit{Phys. Usp.}, 14: 715. \newline

\noindent Vinograd, J., J. Lebowitz and R. Watson. 1968. Early and late
helix-coil transitions in closed circular DNA - number of superhelical turns
in polyoma DNA. \textit{J. Mol. Biol.} 33: 173. \newline

\noindent Volkenshtein, M. V. \emph{Molecular Biophysics} (Acad. Press, New
York, 1977). \newline

\noindent Vologodskii, A. V., A. V. Lukashin, V. V. Anshelevich, and M. D.
Frank-Kamenetskii. 1979. Fluctuations in superhelical DNA. \textit{Nucleic
Acids Res.} 6: 967-982.\newline

\noindent Vologodskii, A. V. and M. D. Frank-Kamenetskii. 1981. Premelting
of superhelical DNA - an expression for superhelical energy. \textit{FEBS
Lett.} 131: 178-180. \newline

\noindent Vologodskii, A. \emph{Topology and Physics of Circular DNA} (CRC
Press Inc., N.W. Boca Raton, Florida, 1992). \newline

\noindent Wada, A. and A. Suyama. 1986. Local stability of DNA and RNA
secondary structure and its relation to biological functions. \textit{Prog.
Biophys. Mol. Biol.} 47: 113-157. \newline

\noindent Wartell, R. M. and A. S. Benight. 1985. Thermal denaturation of
DNA molecules - A comparison of theory and experiment. \textit{Phys. Rep.}
126(2): 67-107.\newline

\noindent Weil , R. and J. Vinograd. 1963. Cyclic helix and cyclic forms of
polyoma viral DNA. \textit{Proc. Natl. Acad. Sci. U.S.A.} 50: 730.



\newpage \centerline{Figure Caption} \noindent{}FIGURE 1 ~ Schematic diagram
of closed circular DNA. \vskip 5 mm

\noindent{}FIGURE 2 ~ Schematic diagram for construction of Hamiltonian for
linear DNA. \vskip 5 mm

\noindent{}FIGURE 3 ~ Tramsformation of the melting curve of linear DNA
(left) into the melting curve of ccDNA (right). The horizontal axis
corresponds to the temperature ($T^\circ K$). The vertical axis shows the
fraction of the open bonds (dimensionless). \vskip 5 mm

\noindent {}FIGURE 4 ~ (a) The scheme of transformation of melting curve for
infinitely sharp helix-coil transition. Along the horizontal axis is the
temperature $(T^{\circ }K)$. The vertical axis shows the fraction of the
open bonds (dimensionless). (b) Trnsformation of the differential melting
curve for infinitely sharp helix-coil transition. The vertical axiss shows
the derivative of the fraction of the open bonds, in units $1/T$. \vskip5
mm\newpage

\end{document}